\documentclass{elsart}
\usepackage{graphicx}

\def\gsim{\;\raise0.3ex\hbox{$>$\kern-0.75em\raise-1.1ex\hbox{$\sim$}}\;}
\def\lsim{\;\raise0.3ex\hbox{$<$\kern-0.75em\raise-1.1ex\hbox{$\sim$}}\;}
\begin{document}

\begin{frontmatter}
{\hfill \small DSF-16/2005, IFIC/05-17, MPP-2005-36}

\title{Relic neutrino decoupling including\\ flavour oscillations}
\author[Napoli,Syracuse]{Gianpiero Mangano},
\author[Napoli]{Gennaro Miele},
\author[IFIC]{Sergio Pastor},
\author[IFIC]{Teguayco Pinto},
\author[Napoli]{Ofelia Pisanti},
\author[MPI]{Pasquale D.\ Serpico}
\address[Napoli]{Dipartimento di Scienze Fisiche, Universit\`{a} di Napoli
{Federico II} and INFN, Sezione di Napoli, Complesso Universitario di
Monte S. Angelo, Via Cintia, I-80126 Naples, Italy}
\address[Syracuse]{Department of Physics, Syracuse University, Syracuse NY,
13244-1130, USA}
\address[IFIC]{Instituto de F\'{\i}sica Corpuscular (CSIC-Universitat de
Val\`{e}ncia), Ed.\ Institutos de Investigaci\'{o}n, Apdo.\ 22085,
E-46071 Valencia, Spain}
\address[MPI]{Max-Planck-Institut f\"{u}r Physik
(Werner-Heisenberg-Institut), F\"{o}hringer Ring 6, D-80805 Munich,
Germany}

\begin{abstract}
In the early universe, neutrinos are slightly coupled when
electron-positron pairs annihilate transferring their entropy to
photons.  This process originates non-thermal distortions on the
neutrino spectra which depend on neutrino flavour, larger for $\nu_e$
than for $\nu_\mu$ or $\nu_\tau$.  We study the effect of
three-neutrino flavour oscillations on the process of neutrino
decoupling by solving the momen\-tum-depen\-dent kinetic equations for
the neutrino spectra. We find that oscillations do not essentially
modify the total change in the neutrino energy density, giving $N_{\rm
eff}=3.046$ in terms of the effective number of neutrinos, while the
small effect over the production of primordial $^4$He is increased by
${\mathcal O}(20\%)$, up to $2.1 {\times} 10^{-4}$. These results are
stable within the presently favoured region of neutrino mixing
parameters.

\end{abstract}
\end{frontmatter}

\section{Introduction}
\label{sec:introduction}

The existence of a relic sea of neutrinos is a generic prediction of
the standard hot big bang model, in number only slightly below that of
relic photons that constitute the Cosmic Microwave Background
(CMB). The presence of the cosmic neutrino background has been
indirectly established both at MeV temperatures by the accurate
agreement between the calculated and observed primordial abundances of
light elements from Big Bang Nucleosynthesis (BBN) and at later epochs
by data on the anisotropies of the CMB and the distribution of Large
Scale Structures (LSS) in the universe, through the contribution of
neutrinos to the radiation content.

These cosmic neutrinos were kept in equilibrium by frequent weak
interactions with other particles until the temperature of the
Universe was of order 2-5 MeV, when these interactions became
ineffective and the process of neutrino decoupling took place. The
standard picture in the instantaneous decoupling limit is very simple
(see e.g.\ \cite{kt}): coupled neutrinos had a momentum spectrum with
an equilibrium Fermi-Dirac (FD) form with temperature $T$,
\begin{equation}
f_{\rm eq}(p)=\left
[\exp\left(\frac{p-\mu_\nu}{T}\right)+1\right]^{-1}\,,
\label{FD}
\end{equation}
which is preserved after decoupling, since both neutrino momenta and
temperature redshift identically with the universe expansion. Here we have
included a potential neutrino chemical potential $\mu_\nu$ that would exist
in the presence of a neutrino-antineutrino asymmetry, but it was shown in
\cite{Dolgov:2002ab} that the stringent BBN bounds on $\mu_{\nu_e}$ apply
to all flavours, since neutrino oscillations lead to flavour equilibrium
before BBN. Thus the contribution of a relic neutrino asymmetry can be
safely ignored.

Shortly after neutrino decoupling the photon temperature drops below the
electron mass, favouring $e^{\pm}$ annihilations that heat the photons. If one
assumes that this entropy transfer did not affect the neutrinos because
they were already completely decoupled, it is easy to calculate the
difference between the temperatures of relic photons and neutrinos
$T_\gamma/T_\nu=(11/4)^{1/3}\simeq 1.40102$. However, the processes of
neutrino decoupling and $e^{\pm}$ annihilations are sufficiently close in time
so that some relic interactions between $e^{\pm}$ and neutrinos exist. These
relic processes are more efficient for larger neutrino energies, leading to
non-thermal distortions in the neutrino spectra and a slightly smaller
increase of the comoving photon temperature, as noted in previous works
(for early references, see \cite{Dicus:1982bz} and the full list given in
the review \cite{Dolgov:2002wy}).

A proper calculation of the process of non-instantaneous neutrino
decoupling demands solving the momentum-dependent Boltzmann
equations for the neutrino spectra, a set of integro-differential
kinetic equations that are difficult to solve numerically. In the
early 1990s several works
\cite{Dodelson:1992km,Dolgov:1992qg,Fields:1992zb} performed
momentum-dependent calculations assuming some approximations, such as
Boltzmann statistics for neutrinos, while the full numerical
computation was later carried out in refs.\
\cite{Hannestad:1995rs,Dolgov:1997mb,Dolgov:1998sf,Esposito:2000hi}.
Finally, a further refinement involves the inclusion of finite
temperature QED corrections to the electromagnetic plasma, as done in
\cite{Fornengo:1997wa,Mangano:2001iu}.

The distortions produced on the neutrino momentum distributions are
small and their direct observation is out of question.  However, they
should be included in a calculation of any observable related to relic
neutrinos. For instance, non-thermal distortions lead to an enhanced
number density of relic neutrinos which modifies e.g.\ the
contribution of massive neutrinos to the present energy density of the
universe. Previous analyses focused on two interesting effects. The
first one concerns the contribution of neutrinos to the total
radiation content of the universe, parametrized in terms of the
effective number of neutrinos \cite{Shvartsman:1969mm,Steigman:1977kc}
$N_{\rm eff}$, through the relation \footnote{This equation only holds
after the reheating is completed, so it would be wrong if used in BBN
calculations. For larger temperatures, one should modify it with
$(4/11)^{4/3}\to (T_{\nu 0}/T_{\gamma 0})^{4}$, where $T_{\nu
0}/T_{\gamma 0}$ traces the evolution derived just by the entropy
conservation law, and add the contribution of $e^\pm$ while not
completely non-relativistic.}
\begin{equation}
\rho_{\rm R} = \left[ 1 + \frac{7}{8} \left( \frac{4}{11}
\right)^{4/3} \, N_{\rm eff} \right] \, \rho_\gamma \,\,,
\label{neff}
\end{equation}
where $\rho_\gamma$ is the energy density of photons, whose value
today is known from the measurement of the CMB temperature. This
equation holds as long as all neutrinos are relativistic and, in
principle, $N_{\rm eff}$ can receive a contribution from other
relativistic relics.  In the following we will restrict our analysis
to the standard case, where the departure of $N_{\rm eff}$ from 3 is
due to neutrino heating by $e^{\pm}$ annihilations. The second effect of
the non-thermal distortions leads to a modification of the outcome of
BBN, in particular a change in the production of primordial $^4$He.
{}From previous works, an increase of order $1.5 {\times} 10^{-4}$ in
the $^4$He mass fraction $Y_p$ was found
\cite{Dodelson:1992km,Dolgov:1992qg,Fields:1992zb,Hannestad:1995rs,Dolgov:1997mb,Dolgov:1998sf,Esposito:2000hi}.
Thus the effect on BBN is small, but it has to be taken into account
in precise BBN codes \cite{Cuoco:2003cu,Serpico:2004gx}.

In general, previous analyses of neutrino decoupling did not include
flavour neutrino oscillations, although their potential effect was already
noted long ago \cite{Langacker:1986jv}. The exception is a work by
Hannestad \cite{Hannestad:2001iy}, who calculated the effect of
two-neutrino oscillations on neutrino heating, finding that the neutrino
energy density was slightly higher while the increase in the primordial
abundance of $^4$He due to non-thermal features of neutrino decoupling
could be even doubled. In this work, some approximations were taken, such
as integrated kinetic equations (quantum rate equations or QREs) and
Maxwell-Boltzmann statistics.  In addition, a very recent paper
\cite{Ichikawa:2005vw} considered the effects of neutrino oscillations in
scenarios with low-reheating temperatures (below 10 MeV), using
momentum-dependent equations with massless $e^{\pm}$ in the collision terms,
but ignoring neutrino-neutrino collisions.

In the present paper we perform a new calculation of neutrino
decoupling solving the momentum-dependent kinetic equations as in
\cite{Dolgov:1997mb,Esposito:2000hi,Mangano:2001iu}, but including
also the effect of flavour neutrino oscillations following the
analysis in \cite{Dolgov:2002ab}.  Our aim is to compare our
results with the simplified analysis in \cite{Hannestad:2001iy},
checking the accuracy of its approximations. The numerical
evaluation of the Boltzmann equations is computationally
demanding, but we do not need to perform a scan over all the space
of neutrino mixing parameters, since presently they are already
known with good precision\footnote{Note that we focus on the
standard case of three active neutrinos and do not consider
active-sterile mixing (see instead e.g.\ refs.\ 
\cite{DiBari:2001ua,Abazajian:2002bj,Kirilova:2002ss,Cirelli:2004cz}).}
(except for the mixing angle $\theta_{13}$ which, as we
will see, has a minor effect). We find that the results concerning
neutrino decoupling remain unchanged within the presently favoured
regions for oscillations parameters.

\section{Neutrino decoupling in presence of flavour oscillations}
\label{sec:oscieffects}

In this section we list the set of equations which rule the process of
neutrino decoupling in the epoch of the early universe prior to BBN. We
describe the main terms that appear in the kinetic equations, in particular
those arising from neutrino oscillations, and describe our method to solve
these equations numerically.

\subsection{Equations}

In order to study neutrino decoupling in the early universe in the presence
of flavour oscillations, we describe the neutrino ensemble in the usual way
by generalized occupation numbers, i.e.\ by $3 {\times} 3$ density matrices for
neutrinos and anti-neutrinos as described in
\cite{Sigl:1993fn,McKellar:1992ja}.  The form of the neutrino density
matrix for a mode with momentum $p$ is
\begin{equation}
\varrho(p,t) = \left (\matrix{\varrho_{ee} & \varrho_{e\mu}&
\varrho_{e\tau}\cr \varrho_{\mu e}& \varrho_{\mu\mu}&
\varrho_{\mu\tau}\cr \varrho_{\tau e} &\varrho_{\tau \mu}&
\varrho_{\tau \tau} }\right). \label{3by3}
\end{equation}
The diagonal elements correspond to the usual occupation numbers
of the different flavours, while the off-diagonal terms are
non-zero in the presence of neutrino mixing. There exists a
corresponding set of equations for the antineutrino density matrix
$\bar{\varrho}$, but in absence of a neutrino asymmetry it is not
needed since antineutrinos follow the same the evolution as
neutrinos.

The equations of motion for the density matrices relevant for our
situation of interest in an expanding universe are~\cite{Sigl:1993fn}
\begin{equation}\label{eq:3by3evol}
i\left(\partial_t-Hp\,\partial_p\right)\varrho_p= \left[ \left(
\frac{M^2}{2p} -\frac{8 \sqrt2 G_{\rm F}\,p}{3 m_{\rm W}^2}{E}
\right),\varrho_p\right] +{C}[\varrho_p]~,
\end{equation}
where $H$ is the Hubble parameter, $G_{\rm F}$ is the Fermi constant
and $m_{\rm W}$ the $W$ boson mass. We use the notation
$\varrho_p=\varrho(p,t)$ and $[{\cdot},{\cdot}]$ denotes the
commutator. The vacuum oscillation term is proportional to $M^2$, the
mass-squared matrix in the flavour basis that is related to the
diagonal one in the mass basis ${\rm diag}(m_1^2,m_2^2,m_3^2)$ via the
neutrino mixing matrix,
\begin{equation}
\left(
    \begin{array}{ccc}
        c_{12} c_{13}
        & s_{12} c_{13}
        & s_{13} \\
        -s_{12} c_{23} - c_{12} s_{23} s_{13}
        & c_{12} c_{23} - s_{12} s_{23} s_{13}
        & s_{23} c_{13} \\
        s_{12} s_{23} - c_{12} c_{23} s_{13}
        & -c_{12} s_{23} - s_{12} c_{23} s_{13}
        & c_{23} c_{13}
\end{array}\right).
\end{equation}
Here $c_{ij}=\cos \theta_{ij}$ and $s_{ij}=\sin \theta_{ij}$ for
$ij=12$, 23, or 13. Since we have assumed CP conservation, there are
five oscillation parameters: $\Delta m^2_{21} = m^2_2 - m^2_1, \Delta
m^2_{31} = m^2_3 - m^2_1, \theta_{12},\theta_{23}$ and $\theta_{13}$.
{}From a global analysis of experimental data on flavour neutrino
oscillations, the values of the first four parameters are determined,
while we only have an upper bound on $\theta_{13}$. As a reference, we
take the best-fit values from ref.\ \cite{Maltoni:2004ei},
\begin{equation}
\left(\frac{\Delta m^2_{21}}{10^{-5}~{\rm eV}^2},
\frac{\Delta m^2_{31}}{10^{-3}~{\rm eV}^2},s_{12}^2,
s_{23}^2,s_{13}^2\right)= (8.1, 2.2, 0.3,0.5,0)\,\, ,
\label{oscpardef}
\end{equation}
while for $\theta_{13}$ we will also consider the value allowed at
$3\sigma$, $s_{13}^2=0.047$.

The decoupling of neutrinos takes place at temperatures of the
order MeV, when neutrinos experience both collisions and
refractive effects from the me\-dium. The latter correspond in
Eq.\ (\ref{eq:3by3evol}) to the term proportional to the diagonal
matrix $E$, that represents the energy densities of charged
leptons. For example, $E_{ee}$ is the energy density of electrons
and positrons. Note that we have neglected two terms in Eq.\
(\ref{eq:3by3evol}) with respect to the complete form shown in
\cite{Sigl:1993fn}. The first one is the usual refractive term
$\sqrt{2}G_{\rm F}L$ that is proportional to the charged-lepton
asymmetries. This asymmetric term is negligible at early times
(high temperatures) compared to the $E$ term, while at
temperatures near $n/p$ freeze out ($T\simeq 1~{\rm MeV}$) it is
negligible compared to the vacuum term $M^2/2p$ for the
mass-squared differences $\Delta m_{31}^2$ and $\Delta m_{21}^2$.
The second refractive term not included in Eq.\
(\ref{eq:3by3evol}) arises from neutrino-neutrino interactions and
is proportional\footnote{ Here the density matrix $\varrho$
$(\bar\varrho)$ is the integrated neutrino (antineutrino) density
matrix so that, for example, $\varrho_{ee}$ is the total number
density of electron neutrinos.}  to $(\varrho-\bar\varrho)$, thus
it vanishes for zero neutrino-antineutrino asymmetry.

Finally, the collisions of neutrinos with $e^{\pm}$ or among
themselves are described by the term $C[{\cdot}]$, which is
proportional to $G_{\rm F}^2$. For the off-diagonal complex terms
of the density matrix we approximate collisions with a simple
damping prescription of the form $C[\varrho_{\alpha\beta}(p)] =
-D_p \varrho_{\alpha\beta}(p)$, with the same damping functions
$D_p$ as in \cite{Dolgov:2002ab}. Instead, for the diagonal ones,
in order to properly calculate the neutrino heating process we
must consider the exact collision integral $I_{\nu_\alpha}$, that
includes all relevant two--body weak reactions of the type $
\nu_\alpha(1) + 2 \rightarrow 3 + 4$ involving neutrinos and
$e^{\pm}$,
\begin{eqnarray}
I_{\nu_\alpha} &\left[ f_{\nu_e},f_{\nu_\mu},f_{\nu_\tau} \right]
= \frac{1}{2\,E_1} \sum_{\rm reactions} \int \frac{d^3
p_2}{2\,E_2\,(2\,\pi)^3} \: \frac{d^3 p_3}{2\,E_3\,(2\,\pi)^3} \:
\frac{d^3 p_4}{2\,E_4\,(2\,\pi)^3} \nonumber \\ &{\times}
(2\,\pi)^4 \, \delta^{(4)} \left( p_1 + p_2 -p_3 - p_4 \right) \,
F \left[ \varrho_{\alpha\alpha}(p_1),f_2,f_3,f_4 \right]
\left|M_{12\rightarrow34}\right|^2 \,\, , \label{Icoll}
\end{eqnarray}
Here $F \equiv f_3f_4\left( 1 - \varrho_{\alpha\alpha}(p_1)
\right)\left( 1 - f_2 \right) - \varrho_{\alpha\alpha}(p_1)f_2\left(
1-f_3 \right)\left( 1-f_4 \right)$ is the statistical factor (when the
particle $i=2,3,4$ is a neutrino $\nu_\beta$ one substitutes $f_i$
with the corresponding diagonal term $\varrho_{\beta\beta}(p_i)$) ,
and $M_{12\rightarrow34}$ is the process amplitude. In ref.\
\cite{Dolgov:1997mb} the complete list of relevant processes and
corresponding squared amplitudes are reported, and it is shown that
some of the integrals can be analytically performed, reducing
$I_{\nu_\alpha}$ to a two-dimensional integral. We have actually
checked that including the effect of mixing in the statistical terms
of Eq.\ (\ref{Icoll}) as in \cite{Sigl:1993fn} leads to very small
modifications of our results (a similar conclusion was found in
\cite{Hannestad:2001iy} with QREs and in the Boltzmann limit).

The kinetic equations for the neutrino density matrix are
supplemented by the continuity equation for the total energy
density $\rho_{\rm R}$,
\begin{equation}
\frac{d\rho_{\rm R}}{dt} =
-3H\,(\rho_{\rm R}+P_{\rm R}) \,\,,
\label{energy}
\end{equation}
where $P_{\rm R}$ is the total pressure of the relativistic
plasma: the three neutrino states and the electromagnetic
components $\gamma$ and $e^{\pm}$ (always in equilibrium with
temperature $T_\gamma$). This equation gives the evolution of the
photon temperature $T_\gamma$. Finally, the finite temperature QED
corrections to the electromagnetic plasma modify the equations of
state of $e^{\pm}$ and $\gamma$ and are taken into account as
described in \cite{Mangano:2001iu}.

\subsection{Computation and technical issues}

Our set of Eqs.\ (\ref{eq:3by3evol}) and (\ref{energy}) are
simplified when we use the following dimensionless variables
instead of time, neutrino momenta and photon temperature:
\begin{equation}
x \equiv mR , \qquad  y \equiv pR , \qquad z \equiv T_\gamma R~,
\end{equation}
where $m$ is an arbitrary mass scale that we choose to be the electron
mass and $R$ is the universe scale factor (normalized so that $R(t)\to
1/T_\gamma$ at large temperatures). For details and the specific forms
of the different terms in the equations, we refer the reader to the
appendix in \cite{Dolgov:2002ab} (but note that in this work the mass
scale $m$ was chosen to be 1 MeV).

The kinetic equations (\ref{eq:3by3evol}) are integro-differential due to
the collision terms in Eq.\ (\ref{Icoll}). In previous works without
neutrino oscillations, the system was solved either using a discretization
in a grid of dimensionless momenta as in
\cite{Hannestad:1995rs,Dolgov:1997mb} or with an expansion of the
non-thermal distortions in moments as in
\cite{Esposito:2000hi,Mangano:2001iu}. Here we follow the discretization
method, calculating the evolution of the neutrino density matrix on a grid
of 100 neutrino momenta in the range $y_i \in [0.02,20]$.

We start to compute the evolution of the system at a value of the
parameter $x_{\rm in}=m_e/(10~{\rm MeV})$, when weak interactions
were effective enough to keep neutrinos in equilibrium with the
electromagnetic plasma. Therefore, the initial
values\footnote{Actually we solve the kinetic equations for the
components of $\varrho(y_i,x)$ normalized to the FD distribution
in Eq.\ (\ref{FD}).}  of the components of the density matrix
$\varrho(y_i,x)$ are either $[\exp(y_i/z_{\rm in})+1]^{-1}$
(diagonal components) or zero (off-diagonal), since flavour
oscillations are suppressed at large temperatures by medium
effects. Finally, the initial value of the dimensionless photon
temperature is $z_{\rm in}=1.00003$, which can be found solving
Eq.\ (\ref{energy}) with neutrinos fully coupled
\cite{Dolgov:1998sf}.

The system of equations is solved from $x_{\rm in}$ until a value of
$x$ when both the neutrino distortions and the comoving photon
temperature $z$ are frozen, approximately at $x_{\rm fin}\simeq
35$. In order to solve simultaneously for the generation of the
distortion and the effect of flavour oscillations we proceed as
follows.  First the change in the diagonal terms of $\varrho(y_i,x)$
from the collision integrals (neutrino heating) is found in a step
$\Delta x$ whose value is set by the typical time scale of
electron-positron annihilation rate (we use 1000 steps in $\log(x)$ in
the range $[x_{\rm in},x_{\rm fin}]$). Then the (fast) effect of the
oscillations is then calculated by following the neutrino density
matrix evolution on a smaller $x$ step, given by $\Delta x$/100.

\section{Results}
\label{sec:results}

We have numerically calculated the evolution of the neutrino
density matrix solving the system of Eqs.\ (\ref{eq:3by3evol}) and
(\ref{energy}), during the full process of neutrino decoupling. In
order to compare with previous results, we have also calculated
the case without neutrino mixing, with and without QED
corrections. When flavour neutrino oscillations are included, we
consider two cases, corresponding to the best-fit values of the
mixing parameters in Eq.\ (\ref{oscpardef}) and either
$\theta_{13}=0$ or $s_{13}^2=0.047$.

\subsection{Evolution of the non-thermal distortions}

\begin{figure}[t]
\includegraphics[width=.95\textwidth]{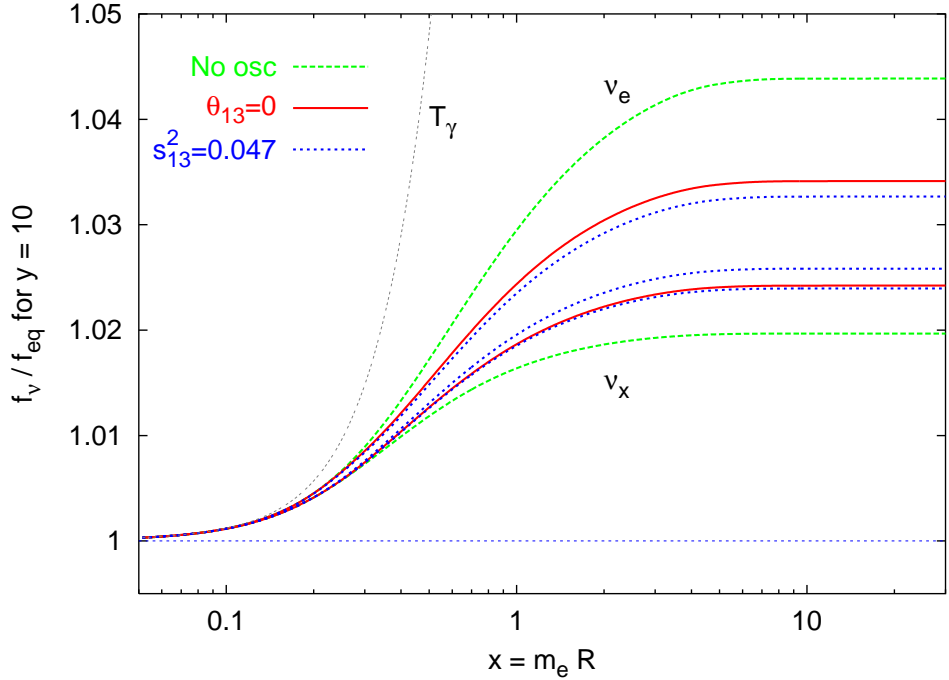}
\caption{\label{fig:evolfnu} Evolution of the distortion of the
$\nu_e$ and $\nu_x=\nu_{\mu,\tau}$ spectrum for a particular comoving
momentum ($y=10$). In the case with $\theta_{13}\neq 0$ one can
distinguish the distortions for $\nu_\mu$ (middle line) and $\nu_\tau$
(lower line). The line labeled with $T_\gamma$ corresponds to the
distribution of a neutrino in full thermal contact with the
electromagnetic plasma.}
\end{figure}

We present in Fig.\ \ref{fig:evolfnu} the evolution of the distortion
of the neutrino distribution as a function of $x$ for a particular
neutrino momentum ($y=10$). In the absence of mixing, the evolution of
$f_\nu$ has been described in previous analyses (see e.g.\
\cite{Dolgov:1997mb,Esposito:2000hi}). At large temperatures or
$x\lsim 0.2$, neutrinos are in good thermal contact with $e^{\pm}$ and
their distributions only change keeping an equilibrium shape with the
photon temperature $[\exp(y/z(x))+1]^{-1}$ (the $T_\gamma$ line in the
figure) . In the intermediate region $0.2 \lsim x \lsim 4$, weak
interactions become less effective in a momentum-dependent way,
leading to distortions in the neutrino spectra which are larger for
$\nu_e$'s than for the other flavours.  Finally, at larger values of
$x$ neutrino decoupling is complete and the distortions reach their
asymptotic values. For the particular neutrino momentum in Fig.\
\ref{fig:evolfnu}, the final value of the distribution is $4.4\%$
($\nu_e$) and $2\%$ ($\nu_{\mu,\tau}$) larger than in the
instantaneous decoupling limit.

It is obvious that flavour neutrino oscillations will modify the
generation of neutrino distortions if they are effective at the
relevant range of temperatures. This depends on the different terms in
the kinetic equations for the neutrino density matrix, in particular
the relative importance of the oscillation term (of order $\Delta
m^2/2p$) which grows as $x^2$, with respect to the background
potential proportional to the energy density of electrons and
positrons (decreases as $x^4$), since the other charged leptons have
already disappeared. In the range $x\lsim 0.3$ the refractive term
dominates, suppressing flavour oscillations so that the neutrino
distributions grow as in the absence of mixing. Then the $e^{\pm}$
potential adiabatically disappears, leading to the usual MSW-type
evolution and a convergence of the flavour neutrino
distortions. Finally, the oscillation term dominates and oscillations
proceed as in vacuum (in an expanding universe, as calculated in
\cite{Kirilova:1988be}). As can be seen for example in Fig.\
\ref{fig:evolfnu}, if $\theta_{13}= 0$ the final value of the
distribution of $\nu_e$'s at $y=10$ is reduced to $3.4\%$ while for
$\nu_{\mu,\tau}$ increases to $2.4\%$. When we take $s_{13}^2=0.047$,
we find $3.2\%$ for $\nu_e$'s and a different distortion for
$\nu_\mu$'s ($2.6\%$) and $\nu_\tau$'s ($2.4\%$).

For a more detailed description of the evolution of flavour
oscillations at this epoch, we refer the reader to
\cite{Dolgov:2002ab}.

\subsection{Frozen spectra and $N_{\rm eff}$}
\begin{figure}[t]
\includegraphics[width=.95\textwidth]{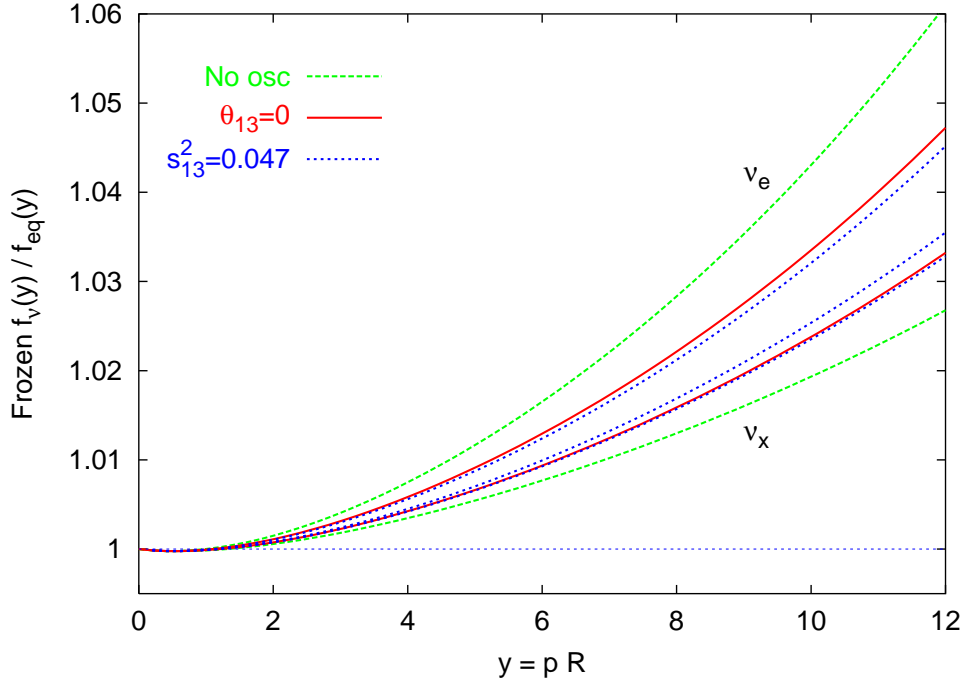}
\caption{\label{fig:finalfnu} Frozen distortions of the flavour
neutrino spectra as a function of the comoving momentum. In the case
with $\theta_{13}\neq 0$ one can distinguish the distortions for
$\nu_\mu$ (middle line) and $\nu_\tau$ (lower line).}
\end{figure}

We show in Fig.\ \ref{fig:finalfnu} the asymptotic values of the
flavour neutrino distribution, for the cases without oscillations and
with non-zero mixing. The dependence of the non-thermal distortions in
momentum is well visible, which reflects the fact that more energetic
neutrinos were interacting with $e^{\pm}$ for a longer period. Moreover,
the effect of neutrino oscillations is evident, reducing the
difference between the flavour neutrino distortions.

Once we have found the final neutrino distributions, the frozen values
of some quantities characterizing neutrino heating can be
calculated. In Tables 1 and 2 we present our results for the
dimensionless photon temperature $z_{\rm fin}$, the change in the
neutrino energy densities with respect to $\rho_{\nu_0}$ (the energy
density in the instantaneous decoupling limit) and the asymptotic
effective number of neutrinos $N_{\rm eff}$ as defined in Eq.\
(\ref{neff}), that can be calculated as
\begin{equation}
N_{\rm eff} = \left( \frac{z_{0}}{z_{\rm fin}}\right)^4 \left( 3
+ \frac{\delta \rho_{\nu_e}}{\rho_{\nu_0}} +  \frac{\delta
\rho_{\nu_\mu}}{\rho_{\nu_0}}  +  \frac{\delta
\rho_{\nu_\tau}}{\rho_{\nu_0}}\right) \,\, ,
\label{neff-emp}
\end{equation}
where $z_{0}=(11/4)^{1/3}\simeq 1.40102$. Note that one can not relate
$z_{\rm fin}$ with a ratio of photon and neutrino temperatures, since
the neutrino spectra are non-thermal and strictly speaking $T_\nu$ is
not defined.

In the absence of mixing, our results in Table 1 without QED
corrections agree with previous works
\cite{Dolgov:1997mb,Esposito:2000hi}, while including QED corrections
we find a slightly smaller $z_{\rm fin}$ than \cite{Mangano:2001iu}
that is due to a more accurate numerical calculation of the evolution
of $z(x)$. After evaluating the precision in the
numerical calculations (modifying our choice for the grid in neutrino
momenta, the initial value of $x$, etc), we estimate that the accuracy
in the values of $N_{\rm eff}$ is $\pm 0.002$. For comparison, we
also show the results of two {\em toy} cases where all neutrino
flavours have the same interactions with $e^+-e^-$ as $\nu_e$'s or
$\nu_\mu$'s, respectively.

When flavour oscillations are taken into account, our results in Table
2 show that, while the modifications in the individual values of
$\rho_{\nu_\alpha}$ can be clearly seen, the contribution of neutrino
heating to the total relativistic energy density is almost unchanged,
with a value of $N_{\rm eff} = 3.046$. The difference with respect to
the unmixed case is only seen in the results within the following
decimal place: $3.0458$ (no oscillations) to $3.0455$ (with
oscillations, either $s^2_{13}=0$ or $0.047$). We checked that even in
the case of bimaximal mixing where $\theta_{12}=\theta_{23}=\pi/4$ and
$\theta_{13}=0$ (disfavoured by present experimental data) the change
in $N_{\rm eff}$ is minimal, with a very small decrease to
$3.0454$. Thus the presence of neutrino oscillations leads to slightly
less efficient neutrino heating\footnote{An analysis of neutrino
heating in presence of non-standard neutrino interactions is presently
under study \cite{Tegua2}.}. Our findings are therefore quite
different than those presented in \cite{Hannestad:2001iy}, where a
very small but positive change in the neutrino energy density with
respect to the unmixed case was found. This difference is probably due
to the approximations used in that paper.
\begin{table*}
\begin{center}
\caption{Frozen values of $z_{\rm fin}$, the neutrino energy densities
$\delta\bar{\rho}_{\nu_\alpha}\equiv
\delta\rho_{\nu_\alpha}/\rho_{\nu_0}$, $N_{\rm eff}$ and $\Delta Y_p$
in the absence of flavour neutrino mixing.}
\begin{tabular}{lccccc}
\hline Case & $z_{\rm fin}$ &
$\delta\bar{\rho}_{\nu_e}$ &
$\delta\bar{\rho}_{\nu_{\mu,\tau}}$ &
$N_{\rm eff}$  & $\Delta Y_p$\\ \hline
No mixing & 1.3978 & 0.94\% & 0.43\% & 3.046&$ 1.71{\times} 10^{-4}$\\
\hline
No mixing (no QED) & 1.3990 & 0.95\% & 0.43\% & 3.035 &1.47${\times} 10^{-4}$\\ No
mixing (all $\nu_e$) & 1.3966 & 0.95\% & 0.95\% & 3.066&$3.57{\times} 10^{-4}$\\
No mixing (all $\nu_\mu$) & 1.3986 & 0.35\% & 0.35\% & 3.031&$1.35{\times}
10^{-4}$\\
\hline
\end{tabular}
\end{center}
\label{tab:unmixed}
\end{table*}

\begin{table*}
\begin{center}
\caption{Frozen values of $z_{\rm fin}$, the neutrino energy densities
$\delta\bar{\rho}_{\nu_\alpha}\equiv
\delta\rho_{\nu_\alpha}/\rho_{\nu_0}$, $N_{\rm eff}$ and $\Delta Y_p$
including flavour neutrino oscillations.}
\begin{tabular}{lcccccc}
\hline
Case & $z_{\rm fin}$ &
$\delta\bar{\rho}_{\nu_e}$ &
$\delta\bar{\rho}_{\nu_\mu}$ &
$\delta\bar{\rho}_{\nu_\tau}$ &
$N_{\rm eff}$ & $\Delta Y_p$\\ \hline
$\theta_{13}=0$ & 1.3978 & 0.73\% & 0.52\% & 0.52\% & 3.046
&$2.07 {\times} 10^{-4}$\\
$\sin^2\theta_{13}=0.047$
& 1.3978 & 0.70\% & 0.56\% & 0.52\% & 3.046
&$2.12 {\times} 10^{-4}$\\
\hline
Bimaximal ($\theta_{13}=0$)
& 1.3978 & 0.69\% & 0.54\% & 0.54\% & 3.045
&$2.13 {\times} 10^{-4}$\\
\hline
\end{tabular}
\end{center}
\label{tab:mixed}
\end{table*}

The effect of neutrino heating on any quantity that
characterizes relic neutrinos is found replacing the Fermi-Dirac
distribution with the spectra as given in Fig.\ \ref{fig:finalfnu}.
Only when neutrinos are still relativistic one finds an integrated
effect of the distortion. For instance, the contribution of
relativistic relic neutrinos to the total energy density is taken into
account just by using $N_{\rm eff} = 3.046$. But in general, for
numerical calculations such as those done by the codes CMBFAST
\cite{Seljak:1996is} or CAMB \cite{Lewis:1999bs}, one must include the
distortions as a function of neutrino momenta.  Our results for the
case with flavour oscillations and $\theta_{13}=0$ (the red lines in
Fig.\ \ref{fig:finalfnu}) are very well reproduced by the analytical
fits
\begin{eqnarray}
f_{\nu_e}(y) & = & f_{\rm eq}(y)
\left[ 1 + 10^{-4}\left(1 - 2.2\, y + 4.1\, y^2 
- 0.047\, y^3\right)\right ]
\nonumber\\
f_{\nu_{\mu,\tau}}(y) & = & f_{\rm eq}(y)
\left[ 1 + 10^{-4}\left(-4 + 2.1\, y + 2.4\, y^2 
- 0.019\, y^3\right)\right ]
\label{fit_fnualpha}
\end{eqnarray}
Note, however, that for any cosmological epoch when neutrino masses
can be relevant one must consider the neutrino mass eigenstates
$\nu_{1,2,3}$. The corresponding momentum distributions can be easily
found from the flavour ones through the relation
\begin{equation}
f_{\nu_i}(y) = \sum_{\alpha=e, \mu, \tau} 
|U_{\alpha i}|^2 f_{\nu_\alpha}(y)
\label{fnumass}
\end{equation}
which, in the case with oscillations and $\theta_{13}=0$ gives the
simple relations
\begin{eqnarray}
f_{\nu_1}(y) &=& 0.7f_{\nu_e}(y)+0.3f_{\nu_x}(y)\nonumber\\
f_{\nu_2}(y) &=& 0.3f_{\nu_e}(y)+0.7f_{\nu_x}(y)\nonumber\\
f_{\nu_3}(y) &=& f_{\nu_x}(y)
\label{fit_fnumass}
\end{eqnarray}
where we have used that $f_{\nu_x}=f_{\nu_\mu}=f_{\nu_\tau}$.

Finally, let us consider the contribution of massive neutrinos to the
present value of the energy density of the Universe. In general, this
must be numerically evaluated for any choice of neutrino masses
$(m_1,m_2,m_3)$ using the distorted distributions described above.
However, in the particular case when neutrino masses are almost
degenerate it is easy to find, using the expressions in Eqs.\
(\ref{fit_fnualpha}) or (\ref{fit_fnumass}), that the contribution of
neutrinos in units of the critical value of the energy density is
\begin{equation}
\Omega_{\nu} = \frac{\rho_\nu}{\rho_c} =
\frac{3m_0}{93.14\,h^2~{\rm eV}}
\label{omeganu}
\end{equation}
where $h$ is the present value of the Hubble parameter in units of
$100$ km s$^{-1}$ Mpc$^{-1}$ and $m_0$ is the neutrino mass
scale. Here the number in the denominator is slightly smaller than in
the instantaneous decoupling limit ($94.12$).

\subsection{Primordial Nucleosynthesis}

Let us now discuss the effects of neutrino heating on BBN, and in
particular on the production of primordial $^4$He. Neglecting neutrino
oscillations, it is well known that the non-thermal neutrino
distortions change the prediction of the primordial $^4$He mass
fraction $Y_p$ by a small amount, of the order
\cite{Dodelson:1992km,Dolgov:1992qg,Fields:1992zb,Hannestad:1995rs,Dolgov:1997mb,Dolgov:1998sf,Esposito:2000hi}
\begin{equation}
\Delta Y_p\simeq1.5 {\times} 10^{-4} \,\, .\label{standardDY}
\end{equation}
One would then naively guess that neutrino oscillations, as a
sub-leading modification, can only marginally change this effect.
However, the small number in Eq.~(\ref{standardDY}) comes out from
subtle cancellations of much larger effects, each one responsible
of changes in $Y_p$ of ${\mathcal O}(10^{-3})$ (see e.g.\
\cite{Dodelson:1992km}). If neutrino oscillations break these
accidental cancellations, the effect could be comparable with that
of Eq.~(\ref{standardDY}). Indeed, this seems to be suggested by
the approximate analysis performed in \cite{Hannestad:2001iy},
where effects up to $\delta Y_p\approx (1.1-1.3) {\times} 10^{-4}$
were found (here we use $\Delta Y_p$ for the total change due to
neutrino heating and $\delta Y_p$ to the {\em net} effect induced
by neutrino oscillations).

In view of the results discussed in the previous sections,
however, where appreciable differences with respect to the picture
presented in \cite{Hannestad:2001iy} are found, the
oscillation-induced $\delta Y_p$ is likely to change. Before
giving our numerical results, we give a simple estimate of the
expected effects. According to the arguments of
\cite{Dodelson:1992km,Fields:1992zb}, in the approximation of {\em
thermal-equivalent} distortions one can perform a perturbative
estimate of the effects of neutrino reheating on the final neutron
fraction $X_n$.  In a framework where the scale factor
$R=T_{\nu_0}^{-1}$ is kept fixed (i.e.\ unperturbed), one can
identify three changes to $X_n$:
\begin{itemize}
\item[(1)] a change in the weak rates due to the distortion to
$\nu_e-\bar{\nu}_e$ spectra, given by
\begin{equation}
\delta X_n^{\nu}\approx -0.1\frac{\delta
T_{\nu_e}}{T_{\nu_e}}\approx -\frac{0.1}{4}\frac{\delta
\rho_{\nu_e}}{\rho_{\nu_e}} \,\, ; \label{deltarhoe}
\end{equation}

\item[(2)] since the energy must be conserved, the overall extra
energy density $\delta\rho_\nu$ is compensated by a decrease of
the electromagnetic plasma contribution
$\delta\rho_{\rm e.m.}=-\delta\rho_\nu$ with respect to the
instantaneous decoupling case. This changes in turn the weak rates
(where the $e^{\pm}$ distributions enter), finally producing
\begin{equation}
\delta X_n^{\rm e.m.}\approx -0.1\frac{\delta
T_\gamma}{T_\gamma}\approx +\frac{0.1}{4}\frac{\delta
\rho_{\nu}}{\rho_{\nu}} \,\, ;\label{deltarhotot}
\end{equation}

\item[(3)] given the high photon entropy, the BBN can start via
the deuterium production $p+n\to \gamma+ d$ only when the universe
has cooled down to a temperature $T_{\rm BBN}\approx 0.07$ MeV. At
this point, the decay of free neutrons (practically the only weak
process since the $n-p$ freezing at $T_{\rm F}\approx 0.7$ MeV) stops
and most of the neutrons are eventually fixed into $^4$He nuclei.
In formulae,
\begin{equation}
X_n(T_{\rm BBN})=X_n(T_{\rm F})\,
e^{-(t(T_{\rm BBN})-t(T_{\rm F}))/\tau_n}\,\,
, \label{ndecay}
\end{equation}
where $\tau_n$ is the neutron lifetime.  The neutrino reheating
changes the time-temperature relationship, thus the
electromagnetic plasma reaches the value $T_{\rm BBN}$ at a
different time given by \cite{Dodelson:1992km}
\begin{equation}
\frac{\delta t_{\rm BBN}}{t_{\rm BBN}}\approx -\frac{\delta
\rho_{\nu}}{\rho_{tot}} \approx -\frac{1}{2}\frac{\delta
\rho_{\nu}}{\rho_{\nu}}.\label{deltatBBN}
\end{equation}
{}From Eqs.\ (\ref{ndecay}) and (\ref{deltatBBN}) one easily
derives the approximate result
\begin{equation}
\delta X_n^{t}\propto \frac{\delta \rho_{\nu}}{\rho_{\nu}} \,\, . \label{deltaXntBBN}
\end{equation}
\end{itemize}
Once using $Y_p\approx 1.33 X_n$ \cite{Dodelson:1992km} and fixing
the constant in Eq.\ (\ref{deltaXntBBN}) in order to get the
result given in \cite{Hannestad:2001iy} for the no-oscillation
case ($\Delta Y_p=1.2 {\times} 10^{-4}$) one can {\em predict} the
value of $\delta Y_p$ starting from $\delta\rho_{\nu_e}$ and
$\delta\rho_{\nu}$.  This simple approach gives $\delta
Y_p^{osc}\approx 1.3 {\times} 10^{-4}$ (and $\delta
Y_p^{Max}\approx 1.6 {\times} 10^{-4}$ for the maximal mixing
case), quite in nice agreement indeed with the numerical findings
of \cite{Hannestad:2001iy} which are respectively $ 1.1$ and $1.3
{\times} 10^{-4}$. This is obviously not unexpected, given the
{\em thermal-equivalent} approximation used.

When applied to our findings, the same formula, with the same
normalization to the data of \cite{Hannestad:2001iy}, would
predict changes of $0.66$ and $0.76 {\times} 10^{-4}$ for the
$\theta_{13}=0$ and $s_{13}^2=0.047$ cases, respectively. The main
conclusion from this simple argument is that the variation $\delta
Y_p$ induced by neutrino oscillations is likely to be smaller than
what found in \cite{Hannestad:2001iy}.

To improve our previous estimates, we have modified the BBN code
developed in the past decade by the Naples group (see refs.\
\cite{Cuoco:2003cu,Serpico:2004gx}).  Notice that $\bar z\equiv
m_e/T_\gamma=x/z$ is used there as independent variable. This
implies that the neutrino heating effects previously described
have to be re-interpreted in such a framework (see Appendix 3 of
ref.\ \cite{Dodelson:1992km} for an account of this issue). By
definition there is no perturbation to the e.m.\ fluid; instead,
the neutrino fluid is not only distorted by reheating, but also
gets a correction from the modified relation $T_{\nu}-T_\gamma$ or
equivalently $R(\bar z)$. Finally, at a fixed $\bar z$, the Hubble
function is obviously altered by the extra energy density due to
neutrino distortions, and the time-temperature equation also gets
a further correction piece which was called $N(\bar z)$ in ref.\
\cite{Serpico:2004gx}. Fixing as a background the spectra where
reheating as well as QED effects were taken into account, we
calculated $\delta Y_p$ as follows. Both $\rho_\nu(\bar z)$,
$N(\bar z)$, and the weak rates in the {\em Born} approximation
were numerically evaluated for the case under consideration and
the {\em background} solution. The differences between the two
cases were treated as perturbations (keeping fixed the $\bar z$ of
{\em background}) and then properly fitted. The use of the Born
rates to calculate these extra terms does not constitute a bad
approximation, since the effects we are dealing with are really
tiny, and higher order corrections as those coming from radiative
processes can be safely neglected.

In Tables 1 and 2 we also include our results for the change in $Y_p$. Our
result for the standard reheating effect, $\Delta Y_p=1.71{\times} 10^{-4}$,
essentially agrees with the evaluation $\Delta Y_p\simeq 1.5{\times} 10^{-4}$
present in literature (see e.g.\ \cite{Fields:1992zb}), where the slightly
larger value arises from a larger $N_{\rm eff}$ from QED corrections. In
\cite{Fields:1992zb} it was noted that the effect on $Y_p$ could be
reproduced by an effective increase $\Delta N_{\rm BBN}$ in the number of
neutrinos of about 0.01 (actually $\simeq$ 0.013, for our findings).
Although academic on the light of present observational accuracies, we warn
the reader that this approximation only works for $^4$He. Indeed for the
other relevant nuclides it produces a change at the ${\mathcal O}(0.1\%)$
level that is exactly in the opposite direction of the true one, as shown
in Table \ref{tab:nuclides}.
\begin{table}
\caption{\label{tab:nuclides} Comparison of the exact BBN results with
the $\Delta N_{\rm BBN}$ approximation.}
\begin{center}
\begin{tabular}{llllllll}
\hline
Nuclide        & Exact          & $\Delta N_{\rm BBN}=0.013$\\
\hline
$\Delta Y_p$   & $1.71{\times} 10^{-4}$ & $1.76{\times} 10^{-4}$\\
\hline
$\Delta$($^2$H/H) & $-0.0068{\times} 10^{-5}$ & $+0.0044{\times} 10^{-5}$\\
\hline
$\Delta$($^3$He/H) & $-0.0011{\times} 10^{-5}$ & $+0.0007{\times} 10^{-5}$\\
\hline
$\Delta$($^7$Li/H) & $+0.0214{\times} 10^{-10}$ & $-0.0058{\times} 10^{-10}$\\
\hline
\end{tabular}
\end{center}
\end{table}

The effects of the oscillations on $\Delta Y_p$ in Table 2 can be
easily explained qualitatively.  A decrease of
$\delta\rho_{\nu_e}/\rho_{\nu_0}$, which is the unavoidable
consequence of neutrino oscillations, leads to an increase in $Y_p$
(see Eq.\ (\ref{deltarhoe})). On the other hand, a decrease of
$\delta\rho_{\nu}/\rho_{\nu_0}$ causes a {\em decrease} of $Y_p$ (see
Eq.\ (\ref{deltarhotot})). This is in fact what we find, so that the
approximate cancellation of the effects (1) and (2) still holds,
differently than what described in \cite{Hannestad:2001iy}, leaving a
sub-leading contribution of the order of few ${\times} 10^{-5}$ as the
effect of neutrino oscillations on $Y_p$.  Modifying the neutrino
mixing parameters only leads to even smaller effects, since $\Delta
Y_p$ changes from $2.07{\times} 10^{-4}$ ($\theta_{13}=0$) to
$2.12{\times} 10^{-4}$ ($s^2_{13}=0.047$). All the previous results
were obtained for a baryon fraction $\omega_b=0.023$, in agreement
with the WMAP determination \cite{Spergel:2003cb}. As already reported
in \cite{Fields:1992zb}, the effects show only a weak dependency on
the exact value of $\omega_b$, for a large interval of values of this
parameter. In particular, in the range $\omega_b=0.020-0.026$ the
changes in the absolute values reported in Tables 1 and 2 are of
${\mathcal O}(1\%)$, while the relative values are practically
unchanged.

In summary, we find that the global change $\Delta Y_p\simeq
2.1{\times} 10^{-4}$ agrees with the results in
\cite{Hannestad:2001iy} because of the inclusion of QED effects,
but the net effect due to oscillations is about a factor 3 smaller
than what previously estimated. We think that the main reason of
the discrepancy is due to the failure of the momentum-averaged
approximation to reproduce the true distortions.

\section{Conclusions}
\label{sec:concl}

In this paper we have performed a new analysis of the effect of
flavour oscillations on the neutrino decoupling phase in the early
Universe. By numerically solving the relevant kinetic equations we
have found the evolution of the distortions on the energy
distributions of neutrinos caused by residual interactions with the
electromagnetic plasma during the electron/positron annihilation
phase.

The inclusion of neutrino oscillations modifies the evolution and
final values of the distortions in the flavour basis, reducing that of
$\nu_e$'s and enhancing those of the other neutrino flavours.  We have
calculated the frozen values of the neutrino distributions, that
should be used in any numerical evaluation of quantities related to
relic neutrinos, like those done in codes such as CMBFAST or CAMB. In
particular, we find that that the asymptotic value of the total
neutrino energy density in presence of flavour oscillations is
essentially unchanged with respect to the unmixed case, being
parametrized by an effective number of neutrinos $N_{\rm eff}=3.046$.
In addition, we found the effect of neutrino heating on the products
of BBN, which in the case of $^4$He is approximately $20\%$ larger
when flavour oscillations are included.

Present bounds on the radiation content of the Universe from CMB and
LSS data, of the order $N_{\rm eff}<7$ (95\% CL), are still far from
the effect caused by neutrino heating. But a value of $\Delta N_{\rm
eff}=0.046$ would be close to the potential sensitivity of
future CMB data from PLANCK, according to the forecast analysis in ref.\
\cite{Lopez:1998aq}. More recent analyses in refs.\
\cite{Bowen:2001in,Bashinsky:2003tk} show that this conclusion was too
optimistic, reducing the sensitivity to $\Delta N_{\rm eff} \sim 0.2$.

Finally, a few words on the detectability of the shift on BBN products
caused by neutrino heating.  For $Y_p$, the effect of the reheating is
below the ${\mathcal O}(0.1\%)$ level, of which only about $0.02\%$
are due to flavour oscillations.  The theoretical uncertainty is at
least of $0.2\%$, while the observational error is likely to be
$\approx 5\%$ and dominated by systematics, with statistical errors at
least at the $1\%$ level (see e.g.~\cite{Serpico:2004gx} and
references therein). Obviously there is no hope to appreciate such
tiny effects: moreover, given the theoretical uncertainties, an {\em
absolute} prediction of the $^4$He yield at the 0.01\% level would
imply a significant improvement in many sub-leading aspects of the BBN
physics and numerics, which is clearly unjustified given the existing
much larger observational uncertainties.  The {\em overall} change due
to the reheating in the other nuclides yields is at most of ${\mathcal
O}(0.1\%)$, and the net effect due to oscillations practically
negligible.  Once considered that the existing (theoretical as well as
observational) uncertainties are not better than ${\mathcal
O}(10-20\%)$ for the case of deuterium, one has to conclude that
definitively one cannot gain information on the {\em standard}
scenario of neutrino oscillations physics from BBN.

\section*{Acknowledgments}

This work was supported by a Spanish-Italian AI, the Spanish grant
BFM2002-00345, as well as a CICYT-INFN agreement. SP was supported
by a Ram\'{o}n y Cajal contract of MEC.  In Munich, this work was
supported in part by the Deut\-sche For\-schungs\-ge\-mein\-schaft
under grant SFB 375.

\end{document}